\documentclass[Times,twocolumn,tighten]{aastex63}

\submitjournal{ApJ}

\shorttitle{Outburst Profiles of GX~339-4}
\shortauthors{R. Bhowmick et al.}

\begin{document}

\title{Relation Between Quiescence and Outbursting Properties of GX 339-4}

\correspondingauthor{Dipak Debnath}
\email{dipakcsp@gmail.com}

\author[0000-0002-7658-0350]{Riya Bhowmick}
\affiliation{Indian Center for Space Physics, 43 Chalantika, Garia St. Road, Kolkata 700084, India}

\author[0000-0003-1856-5504]{Dipak Debnath}
\affiliation{Indian Center for Space Physics, 43 Chalantika, Garia St. Road, Kolkata 700084, India}

\author[0000-0002-6252-3750]{Kaushik Chatterjee}
\affiliation{Indian Center for Space Physics, 43 Chalantika, Garia St. Road, Kolkata 700084, India}

\author[0000-0002-3187-606X]{Shreeram Nagarkoti}
\affiliation{St. Xavier's College, Maitighar, Kathmandu 44600, Nepal}
\affiliation{Indian Center for Space Physics, 43 Chalantika, Garia St. Road, Kolkata 700084, India}

\author[0000-0002-0193-1136]{Sandip Kumar Chakrabarti}
\affiliation{Indian Center for Space Physics, 43 Chalantika, Garia St. Road, Kolkata 700084, India}

\author[0000-0002-8944-9001]{Ritabrata Sarkar}
\affiliation{Indian Center for Space Physics, 43 Chalantika, Garia St. Road, Kolkata 700084, India}

\author[0000-0001-6770-8351]{Debjit Chatterjee}
\affiliation{Indian Center for Space Physics, 43 Chalantika, Garia St. Road, Kolkata 700084, India}
\affiliation{Indian Institute of Astrophysics, Koramangala, Bengaluru, Karnataka 560034}

\author[0000-0001-7500-5752]{Arghajit Jana}
\affiliation{Physical Research Laboratory, Navrangpura, Ahmedabad 380009, India}
\affiliation{Indian Center for Space Physics, 43 Chalantika, Garia St. Road, Kolkata 700084, India}


\begin{abstract}

Galactic black hole candidate (BHC) GX~339-4 underwent several outbursting phases in the past two and a half decades at irregular intervals 
of $2-3$ years. Nature of these outbursts in terms of the duration, number of peaks, maximum peak intensity, etc. varies. We present a 
possible physical reason behind the variation of outbursts. From a physical point of view, if the supply of matter from the companion is
roughly constant, the total energy release in an outburst is expected to be proportional to the quiescence period prior to the outburst 
when the matter is accumulated. We use archival data of RXTE/ASM from January 1996 to June 2011, and MAXI/GSC from August 2009 to 
July 2020 data. Initial five outbursts of GX~339-4 between 1997 and 2011 were observed by ASM and showed a good linear relation between 
the accumulation period and the amount of energy released in each outburst, but the outbursts after 2013 behaved quite differently.
The 2013, $2017-18$, and $2018-19$ outbursts were of short duration, and incomplete or `failed' in nature. We suggest that 
the matter accumulated during the quiescence periods prior to these outbursts were not cleared through accretion due to lack of viscosity. 
The leftover matter was cleared in the immediate next outbursts. Our study thus sheds light on long term accretion dynamics 
in outbursting sources.

\end{abstract}

\keywords{X-Rays:binaries -- stars individual: (GX~339-4) -- stars:black holes -- accretion, accretion disks -- radiation:dynamics}

\section{Introduction}

Stellar mass black hole candidates (BHCs) are mainly two types: transient and persistent. Transient BHCs most of the time stay in the 
`quiescence' phase. Occasionally, these low mass X-ray binaries (LMXBs) become active and trigger an outburst. The only way to detect
an LMXB is to detect the X-ray radiation, coming from the accretion disk of these systems. The disk forms due to accreting matter from the 
companion via Roche-lobe overflow. The nature of two outbursts even for the same black hole (BH) is not similar. In an outburst, 
a rapid evolution of the spectral and temporal properties are observed. Generally, there are four defined spectral states of a black 
hole candidate (BHC): hard (HS), hard-intermediate (HIMS), soft-intermediate (SIMS) and soft (SS) (see, Remillard \& McClintock 2006 for 
a review). Type-I or classical transient BHCs show all four spectral states, forming a hysteresis loop 
(HS$\rightarrow$HIMS$\rightarrow$SIMS$\rightarrow$SS$\rightarrow$SIMS$\rightarrow$HIMS$\rightarrow$HS) during an outburst, whereas 
type-II or harder outbursts do not show softer states(SIMS \& SS) (see, Debnath et al. 2017) during their outbursts. 
There are few exceptions of transient BHCs, such as H~1743-322, present source GX 339-4, showed both types of spectral state evolutions. 
Low frequency quasi-periodic oscillations (LFQPOs) are common features in hard and intermediate spectral states. Sometimes monotonic 
evolution of the LFQPOs is observed during HS and HIMS in both the rising and the declining phases of an outburst. LFQPOs are 
generally observed sporadically in the SIMS (see, Nandi et al. 2012; Debnath et al. 2013 and references therein). These three spectral 
states are also active in jets. Generally, we do not observe any outflows as well as LFQPOs in the SS.

To find a physical explanation of accretion flow dynamics of BHs, many models are put forward in past decades. Two Component Advective 
Flow (TCAF) model is one of such models, was introduced in mid-90s (Chakrabarti \& Titarchuk 1995; Chakrabarti 1997). It is a solution of 
radiative transfer equations considering both heating and cooling effects. According to this model, accretion disk consists of two component 
of flows: a geometrically thin, optically thick, high viscous Keplerian flow and a low viscous, optically thin sub-Keplerian flow or halo. 
The Keplerian flow accretes on the equatorial plane and is immersed within the sub-Keplerian flow. The sub-Keplerian matter moves faster 
than the Keplerian matter. Due to the rise in the centrifugal force close to the black hole, the halo matters temporarily slows down and 
forms an axisymmetric shock at the centrifugal barrier (Chakrabarti 1989, 1990). The post-shock region is a hot and puffed-up region, known 
as CENtrifugal pressure supported BOundary Layer (CENBOL). Multicolor black body spectra are generated from the soft photons originated 
from the Keplerian disk. A fraction of the soft photons from Keplerian disk is intercepted by the CENBOL. They are inverse-Comptonized 
by high energetic `hot' electrons of the CENBOL and produce hard photons. The powerlaw tail in the spectra is produced by these hard photons. 
When the radiative cooling time scale and heating time scale roughly matches, the CENBOL oscillates and the emerging photons produce QPOs 
(Molteni et al. 1996; Ryu et al. 1997; Chakrabarti et al. 2015).

GX~339-4 was first discovered in 1973 by satellite OSO-7 (Markert et al. 1973). The source is located at (l,b)= (338$^{\circ}$.93,-4$^{\circ}$.27) 
with R.A.= $16^h~58.8^m~\pm~0.8^m$ and Dec = $-48^{\circ}41^\prime~\pm~12^\prime$ (Markert et al. 1973). According to Hynes et al. (2004), the 
distance of GX 339-4 is found to be $6\leq d \leq 15$ kpc from the study of high resolution optical spectra of Na~D lines. From optical and 
infrared observations, Zdziarski et al. (2004) estimated the distance to be $d \geq 7$ kpc. Parker et al. (2016) estimated $d = 8.4 \pm 0.9$ 
kpc from the reflection and continuum fitting method using the X-ray spectrum. Since there are no eclipses in the X-ray and the optical data of 
the system, Cowley et al. (2002) suggested that the inclination of the source must lie below $i$ = $60^{\circ}$ and Zdziarski et al. (2004) gave 
an estimation of inclination angle as, $45^{\circ}\leq i \leq80^{\circ}$. Using the relativistic reflection modeling, Parker et al. (2016) found 
an inclination of about $30^{\circ} \pm 1^{\circ}$. Heida et al. (2017) stated that the binary inclination is $37^{\circ} < i < 78^{\circ}$ from 
studies of NIR absorption lines of the donor star. According to Parker et al. (2016) the mass of the source is $9.0_{-1.2}^{+1.6}~M_{\odot}$, 
although Heida et al. (2017) estimated the mass as $2.3~M_{\odot}<~M~<9.5~M_{\odot}$. Recently Sreehari et al. (2019) have evaluated the mass of 
this BH in the range $8.28-11.89~M_{\odot}$ from temporal and spectral analysis. Miller et al. (2008) have suggested the spin parameter of 
the source to be a$=0.93~\pm~0.01$ whereas  Ludlam et al. (2015) found the spin value to be a$>~0.97$ and from relativistic reflection modeling, 
Parker et al. (2016) estimated the spin to be a=$0.95_{-0.08}^{+0.02}$. 

The well known Galactic BHC GX 339-4 is transient in nature, having regular outbursts every $2-3$ year. Since its discovery in 1973, by the 
MIT X-ray detector (on-board the OSO-7 satellite), it was observed by several satellites during different times, such as HAKUCHO, GINGA/ASM, 
LAC, BATSE, SIGMA, RXTE/ASM, MAXI/GSC (Tetarenko et al. 2016). In a period of $46$~years, the total number of outbursts was about $23$.
During the RXTE era (1996 onwards), the source exhibited outbursts in $1997-99$, $2002-03$, $2004-05$, $2006-07$ and $2010-11$ with very low 
luminosity quiescence states in between the outbursts. From 2010 onwards all the outbursts ($2010-11$, $2013$, $2014-15$, $2017-18$, 
$2018-19$ and the latest $2019-20$) are observed by MAXI/GSC.

Although there is debate on the triggering mechanism of an outburst, it is generally believed that an outburst in a black hole candidate is 
triggered due to sudden enhancement of viscosity in the outer edge of the disc (Ebisawa et al. 1996). The declining phase of an 
outburst starts when viscosity becomes weaker. Recently, Chakrabarti et al. 2019 (hereafter CDN19) discussed a possible relation between the 
quiescence phase and the outburst phase of the recurring transient BHC H~1743-322. In this case, matter supplied by the companion starts to 
pile up at a pile-up radius ($X_p$) at the outer disk during the quiescence phase. An outburst could be triggered by a rapid rise of viscosity 
at this temporary reservoir far away from the BH. To find the relation between the outburst and quiescence periods, CDN19 computed the energy 
released during the outbursts of H 1743-322, and showed that on an average, the energy release in an outburst is proportional to the duration 
of the quiescent state (measured as peak to peak flux in between two successive outbursts) just prior to the outburst. Since BHC GX~339-4 also 
underwent several outbursts as the BHC H~1743-322, it would be interesting to check if their conclusion holds good for the present object as 
well.

The paper is organized in the following way. In the next section, we present observation and data analysis methods. In \S 3, we present our 
analysis results. Finally, in \S 4, we discuss our results and make concluding remarks.

\begin{figure*}
\begin{center}
\includegraphics[scale=0.6,width=10truecm,angle=0]{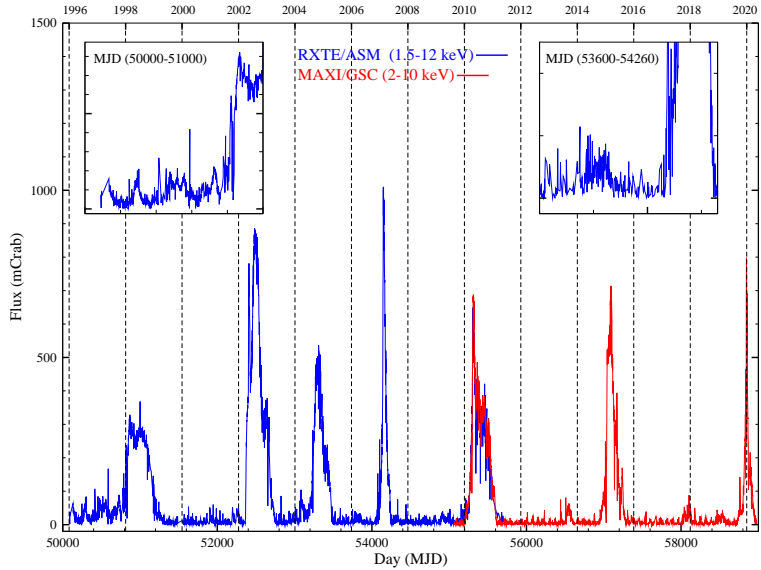} 
\caption{Daily averaged count rate (in mCrab) of RXTE/ASM (blue) in 1.5-12 keV energy band and 
         MAXI/GSC (red) in 2-10 keV energy band.}
\label{fig1}
\end{center}
\end{figure*}

\section{Observation and data analysis}

We use archival data of RXTE/ASM (in $1.5-12$ keV energy range)\footnote{\url{http://xte.mit.edu}} and 
MAXI/GSC (in $2-10$ keV energy range)\footnote{\url{http://maxi.riken.jp}} for our study. 
Our analysis covers ten outbursts of GX~339-4. The daily average light curves are converted into Crab unit using proper conversion 
factors. The Crab conversion factor $75$~Counts/s for the ASM data ($1.5-12$ keV energy range) and $2.82$~photons~cm$^{-2}$~s$^{-1}$ 
for GSC data ($2-10$ keV energy range) are used. For the analysis, we followed the same procedure as described in CDN19.

The nature of the outbursts is different from each other. The light curve of each outburst consists of multiple peaks. To fit each peak 
within an outburst, we use the Fast Rise and Exponential Decay (FRED) profile (Kocevski et al. 2003) as a model written in the ROOT data 
analysis framework of CERN. So to obtain the best fit, we require multiple FRED profiles. The combined FRED fit gives us total energy 
release i.e., the integrated flux of each outburst. As in CDN19, here we also used $12$~mCrab flux value as an outburst threshold to 
calculate the outburst duration. 

According to Nandi et al. (2012 and references therein), one could get a rough idea about the spectral nature of the source from the 
evolution of hardness ratios (HRs; ratio of hard to soft X-ray flux or count rates). GX~339-4 showed all four usual spectral 
states (HS, HIMS, SIMS and SS) during its outbursts except those in $2013$, $2017-18$, and $2018-19$. To calculate the softer 
state (combined SIMS and SS) duration, we used the ratio of $5-12$~keV to $1.5-5$~keV ASM fluxes as HR for RXTE/ASM data and for MAXI/GSC 
HR, we used the ratio of $4-10$~keV to $2-4$~keV fluxes. Low and roughly constant HR values in the middle phase of the outbursts lead 
to the determination of the softer state duration. In case of the `failed' outbursts of $2013$ and $2017-18$, the softer state duration 
dictates the SIMS duration only since in these two outbursts the SS are missing. 
In case of the mini outburst of $2018-19$ both the softer states (SIMS and SS) were missing in the HR diagram.

\section{Results}

A comparative study of the light curves of different sources and the energy release in each outburst helps one to understand the relation 
between the outburst and quiescence phases of black hole X-ray binaries (BHXRBs). Here we study the light curve profiles of the BHC GX 339-4.
We fit the light curve of all the outbursts with multiple FRED profiles. From the FRED fitted curves, we calculate the integrated X-ray flux 
(IFX) in each outburst and make a comparative study. The fits also provide us with peak flux and duration (in MJD) of each outburst. 
Furthermore, we calculate the softer state duration in each outburst from the HR variations and the IFX during softer states (using FRED 
fitted curve). We also examine the relation of peak flux value with softer state IFX and softer state duration. 

\subsection{The Outburst Profile and Integrated Flux Calculation}

Figure 1 shows the RXTE/ASM $1.5-12$~keV light curve (online blue) of GX~339-4 from January 1996 to June 2011 and the MAXI/GSC $2-10$~keV 
light curve (online red) starting from August 2009 to July 2020. The 2013 (F\"{u}rst et al. 2015) and the $2017-18$ 
(Garcia et al. 2019) outbursts were reported as `failed' outbursts, as the source failed to make state transition into the SS. 
The $2018-19$ outburst was also a very short duration `failed' in nature (Paice et al. 2019). Figure 1 shows that the peak flux 
reached maximum value during the $2006-07$ outburst and the peak flux value was minimum during the mini outburst of $2018-19$.

In Fig. 1, we show zoomed pre-outburst periods of the $1997-99$ outburst (MJD=50000 to MJD=51000) in the upper-left corner and 
$2006-07$ outburst (MJD=53600 to MJD=54260) in the upper-right corner. In the case of the $1997-99$ outburst, we can easily see that there are 
multiple small flaring activities before this outburst. Similarly, in $2006-07$ outburst, there is also one significant small flaring 
activity (MJD=53778 to MJD=53886) before the outburst.

As mentioned earlier, we use multiple FRED profiles to fit the RXTE/ASM and MAXI/GSC light curves in all the outbursts of the BHC GX~339-4. 
The shape of the light curve is described by the formula (Kocevski et al. 2003), 
$$F(t) = F_m\left(\frac{t}{t_m}\right)^r\left[\frac{d}{d+r}+\frac{r}{d+r}\left(\frac{t}{t_m}\right)^{r+1}\right]^{-\frac{r+d}{r+1}}, \eqno (1)$$ 
where $F_m$ is the peak value of flux and $t_m$ is the time at which light curve has peak value of flux. Here `r' 
denotes the rising index and `d' denotes the decaying indices.

\begin{figure}
\begin{center}
\includegraphics[scale=0.6,width=10truecm,angle=0]{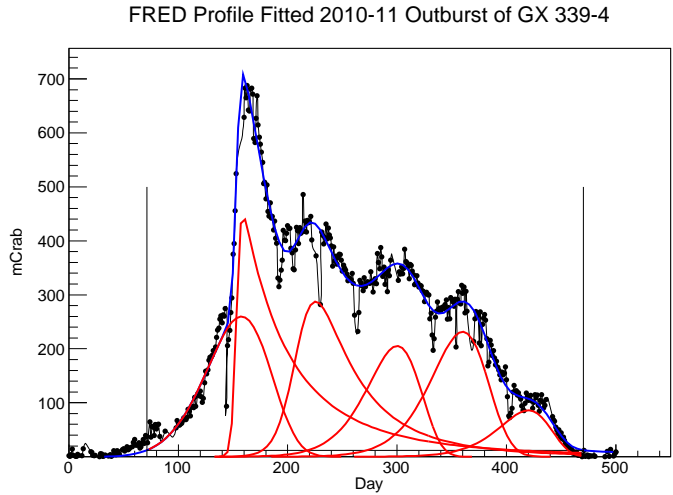} 
\caption{FRED profile fitting on $2010-11$ outburst data. The blue (online) curve is the total FRED fitted curve and the red (online) 
curves are FRED model fitted individual peaks. The flux value of 12mCrab was taken as the quiescent value for all outbursts. A horizontal 
line (black) with value of 12mCrab is drawn. The points where the total FRED-fitted curve (blue) touches the 12mCrab line are the start 
(in rising phase) and end (in declining phase) of the outburst. Beginning  and end  of the outburst are shown as the intersection points 
of the horizontal line with two vertical black lines in the rising and declining phases. Here, MJD=55149.5 is used as 0th day ($t_0$).}
\label{fig2}
\end{center}
\end{figure}

Figure 2 shows the FRED profile (online blue) fitted outburst profile ($2010-11$ in this case), i.e., daily light curve (black lines 
with points). A large fluxctuation in the residual is due to sudden spikes or dips in the data. We use a combination of six FRED profiles 
to fit the profile of the $2010-11$ outburst. To confirm k-number (here six) of FRED profiles (here six) are required for the best fit 
of the multi-peaked outbursts, we used Bayesian Information Criterion (BIC) selection method. $k\pm n$ (n=1,2,..) FRED model fits are 
excluded based on the threshold $\Delta(BIC) > 8$. 

During the quiescent state, the outbursts have a flux equivalent to a few mCrab. A limit of 12 mCrab is used as the threshold for the 
outbursts and energy release above this limit is integrated to measure the integrated flux (IFX) released per outburst. After getting 
the best fit of the outbursts with the FRED model, we mark the start and the end of an outburst based on the 
threshold flux chosen to be 12 mCrab as in CDN19. In Fig. 2, the horizontal line (black) shows the 12 mCrab flux value. The two vertical 
lines (black) mark the start and the end of the outburst. The IFX is calculated from the combined model fit. 
We also calculate various parameters, 
such as peak flux ($F_m$), peak time ($t_m$) and duration of the outburst from the combined FRED fitted curve. The similar fits are done 
for all outbursts and obtained parameters are provided in Table 1. From the $t_m$ values of the outbursts, the periods of accumulation 
of matter from the companion on the disk, as measured from the duration between the peaks of two successive outbursts, are also calculated 
for each outburst.
In Table 2, FRED profile fitted paramters, integral flux and duration of each peak (online red curves) of the outbursts are 
tabulated. The duration and IFX values are calculated within 12 mCrab flux threshold of the individual model fitted curves. 
We also checked wheater the sum of the individual IFXs matches with the total integral flux (online blue curve) of the outburst.

\begin{figure*}
\begin{center}
\includegraphics[scale=0.6,width=12truecm,angle=0]{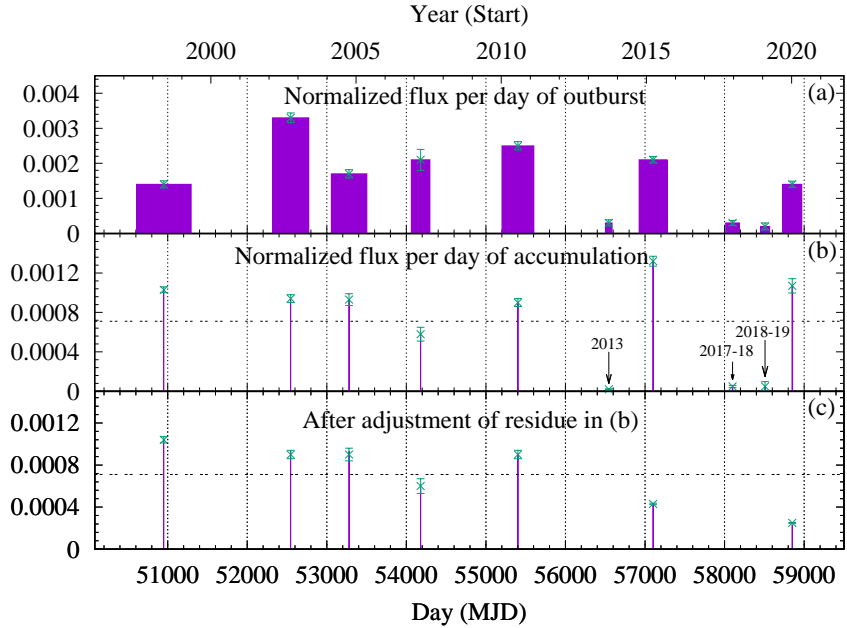} 
\caption{(a) The histogram of the normalized flux per day of each outburst. The width of each histogram shows the duration of the outburst. 
         (b) The normalized flux per day of the accumulation period for each outburst. All integrated fluxes are being normalized using the 
	 $2010-11$ integrated flux value. For the $1997-99$ outburst, the previous outburst is taken in 1995 (Rubin et al. 1998).
         (c) The normalized flux per day of the accumulation period for each outburst where the contribution from 2013 outburst is transferred into the 
	 $2014-15$ outburst, the contribution of the $2017-18$ and $2018-19$ is transferred to the $2019-20$ outburst.}
\label{fig3}
\end{center}
\end{figure*}

From the obtained archival data of RXTE/ASM and MAXI/GSC, the count rates or fluxes and the corresponding uncertainties are 
converted into mCrab unit using conversion factors. We formed three data sets for each outburst: $(i)$ Set-I, MJD vs. flux values, 
$(ii)$ Set-II, MJD vs. error added flux value and $(iii)$ Set-III, MJD vs. error subtracted flux value. We fitted these three data sets 
for each outburst with FRED model and calculated peak flux ($F_m$), peak time ($t_m$), duration, the integrated flux value (IFX), 
the integrated flux value during softer states of outburst from the combined FRED fitted curve. We got the value of the variables 
from each of the three data sets. We calculated the differences of the main value (Set-I) with error added value (Set-II) and main 
value (Set-I) with error subtracted value (Set-III) of each variable obtained from these three data sets. By averaging those, 
we got the `$\pm$' errors in each of the variables.

In Fig. 1, we see that the $2010-11$ outburst was observed by both RXTE/ASM and MAXI/GSC instruments. The values of the variables calculated 
from both RXTE/ASM and MAXI/GSC data are very close to each other (given in Table 1). The value of IFX calculated from RXTE/ASM data 
for the $2010-11$ outburst is 103371 mCrab day and the value of IFX calculated from the MAXI/GSC data is 103829 mCrab day. Due to 
less noise in the GSC data, we have taken 103829 value (the value of IFX of $2010-11$ outburst calculated from MAXI/GSC data) as a 
reference to normalize all the other outbursts.

We observed some peculiarities in the $2013$ and $2014-15$ outbursts. The 2013 outburst has a small peak flux of $\sim51.58$ mCrab and 
minimum duration of $\sim 84$ days. The accumulation period for the 2013 outburst is very high ($\sim 1228$ days) compared to other 
outbursts of GX~339-4. In the immediate next outburst i.e., in $2014-15$, although the accumulation period is small ($\sim 554$ days), it 
showed a higher peak flux ($\sim 619$ mCrab) and a longer outburst duration of $\sim 352$~days. 
The same peculiarity can be seen in case of the $2017-18$, $2018-19$ and the $2019-20$ outbursts. 
In case of the $2017-18$ outburst, there is a large accumulation period of $\sim 1008$~days but the outburst duration is small 
($\sim 173.1$ days) and a small peak flux value of $58.5$~mCrab. Whereas the $2019-20$ outburst has a small accumulation 
period of $\sim314$ days but the outburst has a high peak flux value of $800$ mCrab. Possible reason of this 
behaviour would be discussed in the following sub-Sections.

For each outburst, we also calculated the value of IFX by simply integrating the flux values and calculated the normalized 
IFX w.r.t. the integral flux value of $2010-11$ outburst (given in the Table 1). We noticed that the IFX values from the simple 
integration method does not differ much with that of the FRED fits. But, the use of FRED model is more scientific way to find 
integrated flux and other parameters (peak flux, peak time, duration, etc) of the outbursts. Though, we have downloaded daily 
average ASM and GSC count rate/flux data from the archive, but data for some days are missing. Sometimes there are also sudden rise 
or dip in the light curves for one or two observations. We feel that these are the data errors. So, determination of the parameters 
without fitting with FRED profile will not give the information correctly. Thus, we use the value of the parameters that we get 
from the FRED fitted curve of each outburst for our analysis.

\begin{figure*}
\begin{center}
\includegraphics[scale=0.6,width=11truecm,angle=0]{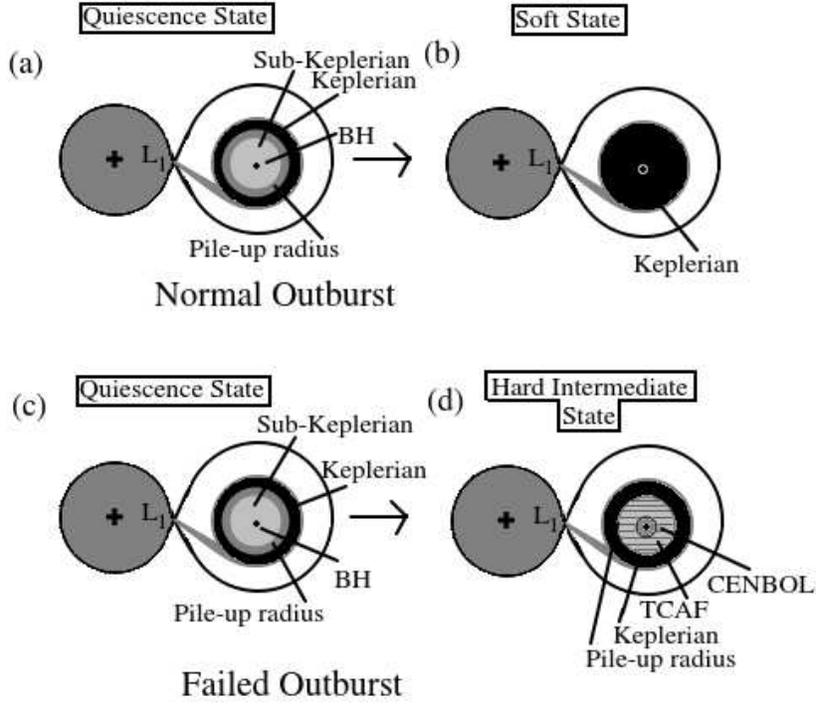} 
\caption{ The possible flow dynamics in case of normal outbursts (a-b) and in case of failed outbursts (c-d). }
\label{fig4}
\end{center}
\end{figure*}

\subsection{A comparative study of outburst fluxes}

The count rates of one day average data from RXTE/ASM and MAXI/GSC are converted to mCrab unit with proper conversion factors (mentioned 
above). Then all the outbursts are fitted with the multiple FRED profiles and the total integrated flux during each outburst is calculated 
in `mCrab day' unit. We sub-divided the IFX of each outburst by the reference value of IFX of the $2010-11$ outburst (observed by MAXI/GSC) 
to get the normalized IFX in each outburst. 

Figure 3(a) shows the IFX per day of outburst, normalized with respect to $2010-11$ outburst (normalized flux per day of an outburst). 
In Fig. 3(a), the area of each histogram represents IFX of each outburst and the width of each histogram represents the duration of each 
outburst. This IFX can be converted into the total energy release rate (Yan \& Yu, 2015) using the prescription provided in http://xte.mit.edu. 

Figure 3(b) shows the normalized IFX per day in the accumulation period of each outburst. For $1997-99$ outburst, the previous outburst 
is taken in 1995 (Rubin et al. 1998) and the peak flux value is taken on MJD=49955. We see that normalized flux per day of accumulation 
is roughly constant except for the outbursts during and after $2013$. In this Figure, we see that the normalized flux per day of accumulation 
is very low at $2013$, $2017-18$ and $2018-19$ outbursts. In case of the $2014-15$ outburst, the normalized flux per day 
of accumulation is the highest (much higher than the average line). In case of the $2019-20$ outburst the value of normalized 
flux per day of accumulation is also high. However, if the supply rate from the companion is almost constant, one might imagine that the 
energy release rate should remain almost the same for all the outbursts.

In order to resolve this anomaly, we may conjecture that perhaps $2013$ outburst did not release its entire accumulated matter and 
the leftover was released in $2014-15$ along with its normal release. This could be tested by adding the energy released in $2013$ and 
$2014-15$ outbursts and distribute the total energy emission rate from 2011 to 2014 evenly. After this exercise, we find that the average 
energy release in $2014-15$ outburst comes down to the average level as par with other outbursts (Fig. 3c).  A similar situation occurred 
during the 2003 outburst of the BHC H~1743-322 as well. Physically, we believe that due to the lack of significant viscosity which triggered 
the 2013 outburst, only a part of the matter accumulated at the pile up radius could accrete. The rest joined with the matter piled up 
subsequently and when the viscosity was high enough, the 2014-2015 outburst started. In CDN19 a general cartoon diagram was presented to 
describe the flow behaviour in H~1743-322 and we believe exactly the same flow dynamics is occurring here as well. 
Just similar to the $2013$ outburst, the $2017-18$ outburst was also `failed' in nature and the leftover matter of these outbursts 
were released in the successive outbursts i.e., in $2018-19$ (mini outburst) and $2019-20$ (main outburst). We combined the energy released 
in these three outbursts and treated them to be the parts of a single outburst (see, Fig. 3c). From the Fig. 3c, it is also evident that 
as time passed, the normalized IFX per day of accumulation was showing a decreasing trend, particular after the $2010-11$ outburst. It 
implies that there could be a non-constant rate of supply of matter from the companion.

In Fig. 4, we show what happens to a failed outburst as opposed to a complete normal outburst. Let us start with the configuration in the 
quiescence state (Fig. 4a). After the matter from the companion crosses the Roche-Lobe and creates an accretion shock, the flow continues 
to move inward till viscosity allows it to remain Keplerian (black annulus). The Keplerian disk halts at the pile-up radius $X_p$ (deep grey 
ring) where matter continues to pile up. Inside this radius, the hot, advective sub-Keplerian matter continues to flow emitting a little X-ray in 
this state. When the piled up matter becomes unstable and possibly larger convective viscosity drains out this accumulated matter completely 
and gradually, the outburst is triggered and reaches the final SS state via HS, HIMS and SIMS state and the Keplerian disk reaches the inner 
stable orbit (Fig. 4b). In a failed outburst, the disk may start in the same way (Fig. 4c). However, a significant viscosity may not be 
sustained for a long time, and the Keplerian disk may not move in all the way to the marginally stable orbit. Only a fraction of the piled-up 
matter will be depleted to produce the TCAF and due to the lower viscosity, the CENBOL does not reach the marginally stable orbit. In that case, 
the outburst will see at most in hard intermediate states or soft intermediate states (Fig. 4d). Due to partial evacuation of matter at pile-up 
radius, the rest of the matter remains there and gets added up with newly supplied matter from the companion which is released in the next outburst.

\subsection{Calculation of softer state duration}

\begin{figure*}
\begin{center}
\includegraphics[scale=0.6,width=10truecm,angle=0]{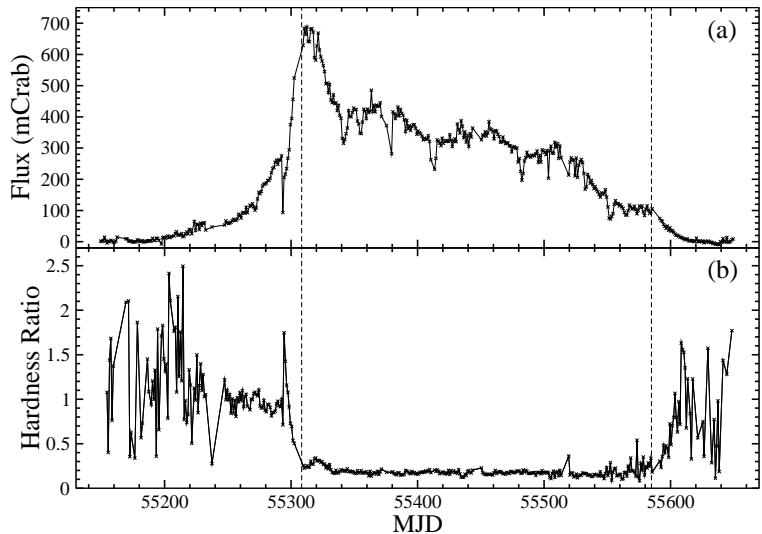} 
\caption{Variation of (a) 2-10 keV MAXI/GSC lightcurve of the 2010-11 outburst and (b) hardness ratio (4-10 keV/2-4 keV count rate) as a function of the 
	Modified Julian Day (MJD). The start and the end of the softer state are shown with two vertical dashed lines.}
\label{fig5}
\end{center}
\end{figure*}

To understand the accretion dynamics and the radiative properties of a BHC, one needs to study both the spectral and temporal properties of 
the source. The variation of the hardness ratio (HR; the ratio between fixed energy hard and soft band photon counts) provides us with a rough 
idea to understand spectral nature and transition dates during an outburst of a transient BHC. Since in HR, we use two fixed energy band 
photon rates/fluxes, transition dates between the spectral states may not be same, i.e., may differ slightly when we confirm from the spectral 
analysis with phenomenological (for e.g., disk black body plus power-law models) or physical models, such as TCAF solution, developed by 
Chakrabarti and his collaborators (Chakrabarti \& Titarchuk 1995; Chakrabarti 1997). TCAF model has successfully designated the spectral 
states of many black hole candidates during their outbursts (Nandi et al. 2012; Debnath et al. 2013, 2014, 2015a,b, 2017, 2020; 
Mondal et al. 2014, 2016; Chatterjee et al. 2016, 2019, 2020; Jana et al. 2016, 2020; Shang et al. 2019). However, according to literature 
(see for example, Debnath et al. 2008; Nandi et al. 2012 and references therein) it is believed that in the harder states HR values are 
higher and in the softer states HR values are low. More distinctly in HS, HR values are roughly constant with high value and in HIMS, HR 
changes rapidly. In the rising HIMS period, HR reduces rapidly and in the declining HIMS, HR increases rapidly. In the softer spectral states 
(SIMS and SS), we see low values of HRs in the middle phase of the outburst, where soft disk black body flux or Keplerian disk photons dominate.

Figure 5(a) shows the 2-10 keV MAXI/GSC lightcurve of $2010-11$ outburst. The corresponding time evolution of the hardness ratio (HR; the ratio 
of photon fluxes in $4-10$~keV and $2-4$~keV bands) is plotted in Fig. 5(b). The origin of the time axis is at 2009 Oct. 26 (MJD=55130). We  
see that HR has a high value at the beginning of the outburst. After some time it gradually decreases. Subsequently, for a comparatively 
long time (from MJD=55308.4 to MJD=55584.9) the value remains almost the same. Then the HR again increases gradually to a high value. 
Variation of the hardness ratio distinctly shows the state transitions from harder to softer spectral states. During the middle region of 
the outburst, low HR values are observed, which corresponds to the softer spectral states. Two dashed vertical lines show the start and 
end of the softer states. For this $2010-11$ outburst, the softer state starts on MJD=55308.4 and ends on MJD=55584.9 with a duration of 
$276.5$~days. A similar analysis is done to calculate softer state durations of all other outbursts of GX~339-4. The results are provided 
in Table 1.

\begin{figure*}
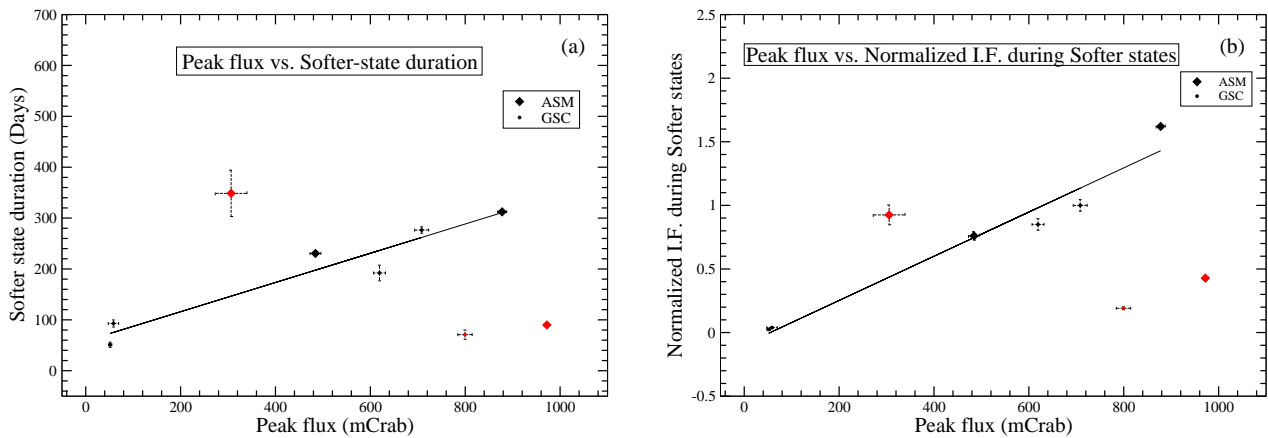

\vskip 0.9cm
\begin{center}
\includegraphics[scale=0.6,width=8truecm,angle=0]{fig6a.eps} 
\hskip 0.65cm
\includegraphics[scale=0.6,width=8truecm,angle=0]{fig6b.eps}
\caption{The correlation in between (a) the peak flux value and the duration of softer states, (b) the peak flux value and normalized 
integrated X-ray flux (IFX) during softer state. In both the figures the RXTE/ASM data points are denoted by diamonds and the MAXI/GSC 
data points are denoted by circles. The outbursts of $1997-99$, $2006-07$ and $2019-20$ are marked as online red since they do not 
follow the general trend as the other outbursts.}
\label{fig6}
\end{center}
\end{figure*}

\vspace{0.5cm}
\subsection{Peak flux relation with softer states}

Here we attempt to find the correlation between the peak flux values ($F_m$) of the outburst, obtained from FRED profile 
fits and softer state durations during the GX~339-4 outbursts. We calculated the peak flux value from the outburst profile of each outburst, 
and the softer state duration from the evolution of the HR. In Fig. 6(a), we plot the peak flux value ($F_m$) vs. softer state duration.
The better the linear regression fits the data, the closer the value of R-square is to $1$ and a correlation is statistically 
significant if the P-value is less than $0.05$ (typically $\leq 0.05$). For the correlation in between $F_m$ and softer state duration the 
values of R-squared and P-value are obtained $0.923$ and $0.002$ respectively. We also calculated the integrated flux during the softer 
state ($IFX_{softer}$) of each outburst and normalized them with respect to the $IFX_{softer}$ of the $2010-11$ outburst. Figure 6(b) shows 
the peak flux value ($F_m$) vs. the normalized IFX in softer states ($IFX_{softer}$). The R-squared value and P-value for the 
correlation in between $F_m$ and $IFX_{softer}$ are 0.960 and 0.0006 respectively. We found linear relations in both the plots 
(Fig. 6a \& 6b). But the outbursts of $1997-99$, $2006-07$ and $2019-20$ do not follow this linearity condition.

The peak flux is decided by the maximum supply of the Keplerian matter, i.e., when the viscosity of the accretion disk is maximum. After that, 
the viscosity is turned off and supply of the disk matter from the pile-up radius is reduced, triggering the onset of the declining phase. 
The larger the value of the peak flux, the longer it takes to wash out the Keplerian matter after the viscosity is turned off 
(Roy \& Chakrabarti 2017). In Roy \& Chakrabarti (2017), theoretically, it is done by mixing Keplerian and sub-Keplerian halo components 
and then draining the mixture. Otherwise, it is difficult to drain the high angular momentum Keplerian matter from the system as the viscosity 
is reduced. Thus, the source remains in the softer states (SS and SIMS) for a long time. So, there is a correlation between the peak flux value 
(maximum viscous day) with the $IFX_{softer}$ and the duration of the softer states. In Figs. 6(a) \& 6(b), we see a general trend of a linear 
relation between these. However, in both the Figures, we see that the $1997-99$, $2006-07$ and $2019-20$ outbursts (marked as online 
red) do not follow the general trend as the other outbursts. When we examine the profiles of the $1997-99$ and $2006-07$ outbursts, we see 
that just before the start of these two outbursts there were many smaller flaring activities, which may have depleted some of the accumulated 
matter at the pile-up radius (see zoomed insets containing the pre-outburst phase in Fig. 1). Multiple small flaring activities before the 
$1997-99$ outburst was reported earlier (Zdziarski et al. 2004). In the case of $2006-07$ outburst, there was a very small outburst during MJD 
$53751-53876$ (Buxton et al. 2012). In prior to the $2019-20$ outburst, there was two `failed' outbursts; the 2017-18 outburst and 
a mini outburst ($2018-19$ outburst). These two outbursts might be the reason for which the $2019-20$ outburst did not follow the 
linear relation as shown in Fig. 6(a) and 6(b).

\section {Discussion and conclusions}

In the post RXTE era (1996 onwards) the source GX~339-4 showed several outbursts at irregular intervals of $2-3$ years. The nature of each of 
these outbursts is different. There are differences in duration, accumulation period (Quiescence phase), maximum peak intensity, etc. in these
outbursts. Recently, Chakrabarti and his collaborators were quite successful to find out the physical reason behind the variation of the nature 
of outbursts of the Galactic transient H~1743-322 (see, CDN19). They found a relation between the energy release at an outburst and
the duration between the current and the previous peak times.

The supply of matter from the companion via Roche-lobe overflow continues in rising as well as the declining and quiescence phases. Generally,
an outburst is triggered when viscosity rises above a critical value and the declining phase starts when the viscous effect is turned off 
(see, Chakrabarti et al. 2005 and references therein). Matter from the outer disk inside the Roche-lobe of the accretor may not be able to 
form a Keplerian disk till the marginally stable orbit due to lack of significant viscosity. Initially, due to low viscosity, this matter 
starts to accumulate at a location, far away from the black hole, known as pile-up radius ($X_p$). With an increase in the amount of 
accumulated matter, the thermal pressure rises. As a result, turbulence and instability increases, which in turn increases the viscosity above
the critical value. With this high viscosity, matter rushes towards the black hole in both Keplerian and sub-Keplerian components (TCAF) from 
$X_p$ and triggers an outburst. While releasing most of the stored hot matter at the pile-up radius ($X_p$), the viscosity is turned off 
triggering an onset of the declining phase followed by the quiescent state. The quiescence phase duration varies with $X_p$. For larger $X_p$, 
the quiescence phase duration is also high because, a larger value of $X_p$ requires higher viscosity to trigger the outburst, i.e., more 
accumulation of matter is required, resulting in larger duration of quiescence. Some examples were given in CDN19. The matter accumulated at 
$X_p$ from the peak day of the previous outburst up to the peak day of the ongoing outburst contributes to the current outburst. Thus the 
total energy released in an outburst reflects the amount of matter accumulated at the pile-up radius during the accumulation period prior to 
that outburst. 

The relation between the outbursts and quiescence periods are also studied for the Galactic transient GX~339-4. To calculate accumulation 
duration of the matter at $X_p$, we considered peak (of the previous outburst) to peak (of the considering outburst) durations, which also include 
the quiescence periods. We notice that the average energy release in an outburst is proportional to the period of accumulation just prior to 
it, with the exception of the outbursts which occurred in $2013$, $2014-15$, $2017-18$, $2018-19$ and $2019-20$. We found 
that the 2013 and $2017-18$ outbursts were `failed' outbursts where the energy released were not complete. If we combine the energies released 
at the 2013 and $2014-15$ outbursts and treat them to be the parts of a single outburst, then the linear relation could be seen. The possible 
flow dynamics in a normal and failed outbursts are shown in Fig. 4. In Fig. 4(c-d), only a part of the matter is released as in 2013 
outburst. The leftover matter at the pile-up radius was combined with freshly supplied matter from the companion and produced the normal 
outburst of $2014-15$ with higher than the expected flux. Similar to the $2013$ outburst, the $2017-18$ is also an incomplete outburst. 
We also found that the $2019-20$ outburst released more than its share as in $2014-15$ outburst. There was also a mini outburst of 
$2018-19$ in between the $2017-18$ and $2019-20$ outbursts. We combined the energies released in the three successive outbursts of 
$2017-18$, $2018-19$ and $2019-20$, and treated them to be the parts of a single outburst that took place in $2019-20$.

We also show the relationship between the peak flux ($F_m$) and softer (SIMS and SS) states. When we plot the peak flux ($F_m$) vs. softer state 
duration (Fig. 6a) and peak flux vs. normalized integrated flux in softer states (Fig. 6b), we find a linear correlation in both the plots. 
The peak flux value indicates the amount of matter accumulated during the rising phase, and which clears out during softer states. So, the higher
the value of the peak flux, the larger is the value of the softer states' duration. Although, in both Figs. 6(a-b), the outbursts 
$1997-99$, $2006-07$ and $2019-20$ do not show similar trends as in other outbursts of GX 339-4. We found a possible reason for this 
anomalous behaviour. It seems that prior to both of these main outbursts, piled up matter was `leaking' and gave rise to some weaker 
flares even in the so-called quiescence phase. In CDN19, they proposed that the 2003 outburst of the BHC H~1743-322 stopped prematurely. 
The following outburst of 2004 released the remaining part of the energy that was due to be released in 2003 outburst. The matter accumulated 
at the pile up radius before the 2003 outburst could not get cleared during the 2003 outburst but remained stuck at a nearer pile up radius. 
After 2003 outburst, more matter accumulated at the new pile up radius and triggered the subsequent 2004 outburst. The duration of the 2003 
outburst was 230.5 days, and the duration of the 2004 outburst was 112.2 days. As the pile up radius prior to 2004 outburst ($X_{p2004}$) was 
closer (to BH) than the pile up radius prior to outburst of 2003 ($X_{p2003}$), the duration of 2004 outburst was smaller than the outburst 
of 2003. In case of the outburst of $2006-07$ and $2019-20$ of the BHC GX~339-4, a small softer state duration was observed (see, Fig. 6). 
During the small `failed' outbursts prior to these two outbursts, matter started to rush from the pile up radius ($X_{pfailed}$) towards 
the black hole but due to sudden fall in viscosity, a large part of the infalling matter stopped prematurely and remained stuck at a new pile up 
radius nearer to the BH. Freshly accreted matter from the companion was also accumulated at the new pile up radius and gave rise to the viscosity. 
When the viscosity raised sufficiently, these accumulated matter at the new pile up radius, started to fall towards the black hole again 
and produced the main outburst. Since the pile up radii in these two cases i.e., prior to the main outbursts of $2006-07$ and $2019-20$ are 
nearer to the BH ($X_{pmain}<X_{pfailed}$), thus the duration of these two outbursts were small. If the pile up radius is nearer to the BH, 
relatively less accumulation of matter is required to trigger the outburst, since the disc is hotter and ionized there. As the duration 
of these two outbursts were small, softer state durations were also shorter. That is why these two outbursts are outlier in Fig. 6a and 6b. 
In case of the $1997-99$ outburst, we are unable to properly predict and discuss the accumulation location and accretion flow as there 
was lack of sufficient data before the start of the outburst. On the eve of this outburst, only some small flaring activities are seen. 
What happened before that is not clear from the data. Beside inward shift of the pile up radius due to the failed outbursting attempts
prior to the main outburst, outburst duration as well as softer state duration and integrated flux during softer state of the present source,
also influenced by the non-constant rate of mass supply from the companion. It is evident from the Fig. 3c that Normalized flux per
day of the accumulation period showed a decreasing trend as time passed, particularly after the 2010-11 outburst. This means that supply rate
from the companion seems to have decreased as the day progressed. In the near future, GX 339-4 may proceed to a long duration quiescence phase.

In any case, our global picture of accumulation of matter at the pile-up radius and its release rate as per availability of viscous processes 
hold good for the present source. This was also seen to be the case in outbursts of H 1743-322. In future we will verify the findings of our work
with other transient BHCs and the results would be reported elsewhere.

\section*{Acknowledgements}

This work made use of ASM data of NASA's RXTE satellite and GSC data provided by RIKEN, JAXA and the MAXI team.
R.B. acknowledges support from CSIR-UGC NET qualified UGC fellowship (June-2018, 527223)
D.D. and S.K.C. acknowledge support from Govt. of West Bengal, India and ISRO sponsored RESPOND project (ISRO/RES/2/418/17-18) fund.
K.C. acknowledges support from DST/INSPIRE (IF170233) fellowship.
D.C. and D.D. acknowledge support from DST/SERB sponsored Extra Mural Research project (EMR/2016/003918) fund.
A.J. and D.D. acknowledge support from DST/GITA sponsored India-Taiwan collaborative project (GITA/DST/TWN/P-76/2017) fund.
A.J. acknowledges the support of the Post-Doctoral Fellowship from Physical Research Laboratory, Ahmedabad, India, funded by the 
Department of Space, Government of India.  

{}

\begin{table}
\vskip 0.5cm
\addtolength{\tabcolsep}{-3.50pt}
\small
\begin{center}
\caption{Peak data and normalized integrated X-ray flux (IFX) during outbursts of GX 339-4}
\label{tab:table1}
\begin{tabular}{|c|c|c|c|c|c|c|c|c|}
\hline
 Outburst & Peak Day & Peak flux  & Duration  & Accumulation$^a$ & Normalized IFX & Normalized IFX$^b$   &Softer state  & Normalized IFX of   \\
   Year   &          &            &           & Period           & w.r.t.         & w.r.t. $2010-11$     & duration     & softer state w.r.t. \\
	  & (MJD)    & (mCrab)    & (Days)    & (Days)           & $2010-11$      &(simple integration)  & (Days)       & $(2010-11)$         \\[0.5ex]
\hline
	RXTE/ASM&&&&&&&&\\
$1997-99$ & 50873.5  & $305.5\pm33.5$  & 686.0 &  918.5 & $0.946\pm0.061$ & $0.951\pm0.094$ & $348.6\pm45.5$   & $0.925\pm0.078$   \\ 
$2002-03$ & 52495.5  & $877.8\pm 9.6$  & 454.3 & 1622.0 & $1.519\pm0.066$ & $1.522\pm0.096$ & $312.3\pm0.90$   & $1.620\pm0.003$   \\
$2004-05$ & 53303.4  & $484.4\pm11.5$  & 446.7 &  806.9 & $0.750\pm0.050$ & $0.757\pm0.072$ & $230.5\pm3.60$   & $0.760\pm0.033$   \\
$2006-07$ & 54152.8  & $972.0\pm 5.5$  & 232.6 &  849.8 & $0.490\pm0.059$ & $0.482\pm0.038$ & $89.90\pm1.50$   & $0.428\pm0.016$   \\  
$2010-11$ & 55308.2  & $689.8\pm10.2$  & 401.9 & 1156.6 & $0.995\pm0.007$ & $0.986\pm0.067$ & $274.2\pm5.12$   & $0.999\pm0.026$   \\  [0.5ex]
\hline                                                                                     
	MAXI/GSC&&&&&&&&\\                                                               
$2010-11$ & 55308.5  & $708.1\pm14.7$  & 399.2 & 1156.3 & $1.000\pm0.051$ & $1.00\pm0.076$  & $276.5\pm6.30$   & $1.000\pm0.046$   \\
2013      & 56536.9  & $51.58\pm 3.5$  &  84.0 & 1228.4 & $0.029\pm0.007$ & $0.028\pm0.006$ & $51.10\pm5.10^*$ & $0.025\pm0.004$   \\
$2014-15$ & 57090.7  & $619.3\pm12.5$  & 352.1 &  553.8 & $0.735\pm0.041$ & $0.737\pm0.069$ & $192.1\pm15.4$   & $0.850\pm0.045$   \\
$2017-18$ & 58098.7  & $ 58.5\pm10.9$  & 173.1 & 1008.0 & $0.047\pm0.012$ & $0.049\pm0.011$ & $92.90\pm6.90^*$ & $0.039\pm0.008$   \\  
$2018-19$ & 58530.5  & $29.43\pm10.8$  & 115.7 &  431.8 & $0.022\pm0.012$ & $0.025\pm0.010$ & $------$         & $------$          \\
$2019-20$ & 58844.1  & $799.5\pm15.1$  & 238.9 &  313.6 & $0.335\pm0.023$ & $0.336\pm0.037$ & $71.0\pm9.00$    & $0.191\pm0.011$   \\  [0.5ex]
\hline
\end{tabular}
\end{center}
\leftline{$^a$ Accumulation period is taken from the peak day of previous outburst to the concerned outburst.}
\leftline{$^*$ In case of $2013$ and $2017-18$ outbursts the softer state duration dictates the SIMS duration only.}
\leftline{$^b$ Integral flux is obtained by simple integration method and then normalized w.r.t. the integral flux of $2010-11$ outburst.}
\end{table}

\begin{table}
\addtolength{\tabcolsep}{-3.5pt}
\centering
	\caption{FRED Profile Fitted Parameters, Integral flux values and Duration in Each Peak of the GX~339-4 Outbursts}
\label{tab:table1}
\begin{tabular}{|c|c|c|c|c|c|c|c|c|c|}
\hline
 Outburst  &  K  &   K1        &    K2      &    K3      &    K4        &   K5       &    K6   &  K's Integral Flux   &  K's Duration  \\ 
   (1)     & (2) &   (3)       &   (4)      &   (5)      &    (6)       &   (7)      &    (8)  &        (9)           &       (10)     \\ [0.6ex]

\hline
 1997-99   &  5  & 35.4, 50673, & 49.6, 50729, & 197, 50867, & 263, 51029,  & 18.9, 51246, & - - - & 6299, 1148, 190003, & 202, 31.0, 227, \\
           &     & 0.755, 5.39  & 15.2, 86.4   & 13.3, 7.46  & 3.48, 8.2E+6 & 378, 10.6    & - - - & 70090, 814          & 516, 35.8       \\[0.3ex]
\hline
 2002-03   &  4  &  291, 52386, & 498, 52411 & 871, 52496, & 213, 52650,   & - - - & - - - & 12623, 10235, & 116, 59, \\
           &     &  43.8, 11.0  & 81.3, 28.8 & 11.6, 8.06  & 23.02, 57.7   & - - - & - - - & 120089, 14760 & 448, 142 \\[0.3ex]
\hline
 2004-05   &  3  & 62.3, 53095, & 481, 53303, & 126, 53434,   & - - - & - - - & - - - & 4755, 63259, 9851 & 115, 339, 139\\ 
           &     & 3.86, 4.20   & 7.68, 10.02 & 13.54, 1.1E+7 & - - - & - - - & - - - &                   &              \\ [0.3ex]
\hline

 2006-07   &  3  & 157, 54124, & 942, 54154, & 19.52, 54220, & - - - & - - - & - - - & 7598, 42395, 867 & 90, 179, 33.9\\
           &     & 6.74, 17592 & 30.9, 7.32  & 13.40, 1.1E+7 & - - - & - - - & - - - &                  &              \\[0.3ex]
\hline

 2010-11   &  6  & 260, 55307, & 446.9, 55309, & 287, 55375, & 205, 55450,  & 232, 55510, & 85.8, 55570, & 19054, 27898, 22757, & 146, 264, 220, \\
           &     & 4.9, 1.1E+6 & 76.8, 3.87    & 20.5, 6.19  & 11.8, 6.9E+6 & 13.9, 202   & 17.5, 1.2E+7 & 13094, 15718, 5308   & 126, 138, 100  \\[0.3ex]
\hline

 2013      &  1  & 51.6, 56536, & - - - & - - - & - - - & - - - & - - - & 2972.92 & 83.98 \\
           &     & 1.17, 2.6E+6 & - - - & - - - & - - - & - - - & - - - &         &       \\[0.3ex]
\hline

 2014-15   &  5  & 86.5, 57005, & 393, 57042, & 488, 57095, & 180, 57163, & 117, 57240, & - - - & 4758, 21833, 37099, & 90.7, 228, 182, \\
           &     & 5.75, 62.6   & 151, 4.2    & 8.92, 20.8  & 40.2, 9.97  & 58.5, 92.9  & - - - & 10105, 2569         & 132, 39.2       \\[0.3ex]

\hline
 2017-18   &  3  & 34.7, 58044, & 56.5, 58100, & 48.5, 58178, & - - - & - - - & - - - & 1804, 2619, 416 & 64.3, 66.5, 12.5\\ 
           &     & 1.6, 1.5E+6  & 10.4, 6.67   & 109, 45.6    & - - - & - - - & - - - &                 &                 \\ [0.3ex]

\hline
 2018-19   &  2  & 20.9, 58478,  & 22.6, 58533, & - - - & - - - & - - - & - - - & 1402, 885 & 63.4, 42.4\\
           &     & 0.935, 4.7E+6 & 15.9, 3.09   & - - - & - - - & - - - & - - - &           &           \\[0.3ex]
\hline

 2019-20   &  5  & 37.4, 58769, & 463, 58848,  & 348, 58844, & 82.8, 58879, & 133, 58902, & - - - & 2475, 20149, 3860, & 83.3, 96.3, 37.8, \\
           &     & 1.62, 34.9   & 7.55, 1.3E+7 & 954, 13.7   & 151, 6.87    & 22.5, 11.2  & - - - & 2582, 5759         & 61.0, 88.1        \\[0.3ex]

\hline
\end{tabular}
\leftline{In Col. 2, K denotes the number of FRED models used to fit the entire outburst.}
\leftline{K1-K6 mark individual model fitted parameters: peak flux ($F_m$ in $mCrab$), peak day ($t_m$ in MJD), rising index ($r$), decaying index ($d$).}
\leftline{Col. 9 \& 10 mark integral flux and duration of the individual peaks in units of $mCrab~day$ and $day$ respectively.}
\end{table}

\end{document}